# Redesigning Computer-Based Learning Environments: Evaluation as Communication

Matthias R. BRUST
Faculty of Sciences, Technology and Communication, University of Luxembourg
L-1359 Luxembourg, Luxembourg

and

Christian M. ADRIANO
Echo Colégio Americano
Santa Maria - RS, Brazil

and

Ivan L. M. RICARTE
Dep. of Computer Engineering and Industrial Automation, State University of Campinas
Campinas - SP, Brazil

## ABSTRACT

In the field of evaluation research, computer scientists live constantly upon dilemmas and conflicting theories. As evaluation is differently perceived and modeled among educational areas, it is not difficult to become trapped in dilemmas, which reflects an epistemological weakness. Additionally, designing and developing a computer-based learning scenario is not an easy task. Advancing further, with end-users probing the system in realistic settings, is even harder. Computer science research in evaluation faces an immense challenge, having to cope with contributions from several conflicting and controversial research fields. We believe that deep changes must be made in our field if we are to advance beyond the CBT (computer-based training) learning model and to build an adequate epistemology for this challenge. The first task is to relocate our field by building upon recent results from philosophy, psychology, social sciences, and engineering. In this article we locate evaluation in respect to communication studies. Evaluation presupposes a definition of goals to be reached, and we suggest that it is, by many means, a silent communication between teacher and student, peers, and institutional entities. If we accept that evaluation can be viewed as set of invisible rules known by nobody, but somehow understood by everybody, we should add anthropological inquiries to our research toolkit. The paper is organized around some elements of the social communication and how they convey new insights to evaluation research for computer and related scientists. We found some technical limitations and offer discussions on how we relate to technology at same time we establish expectancies and perceive others work.

## 1. INTRODUCTION

In the research field of computer-based learning, evaluation is by far the less understood issue. Computer scientists in this field live constantly upon dilemmas and conflicting theories, since evaluation is differently perceived and modeled among educational areas. Should the assessment tools in the learning environment be based on discursive or multiple-choice questions? Should questioning be objective or subjective? Should evaluation be based on reasoning or on hands-on experience? Should it be conducted by the student or by the tutor? Designing and developing a computer-based learning environment is a tough work, but to be trapped in these dilemmas reflects an epistemological weakness. The issue becomes more complex when recognizing that good evaluation requires much more than simple assessment tools, with many events of interest that can occur and be recorded by the computer. Although more information is the basis for a better evaluation, there is no simple answer to support the selection of relevant events and the analysis of acquired data.

Computer science research in evaluation imposes a challenge being among conflicting and controversial research fields – philosophy and psychology, social sciences and engineering. Has computer science an adequate epistemology for such a challenge? Were the CBT (Computer-based Training) systems the far we could go? We believe not, but deep changes must be made in our field. The first task is to relocate this research field by building upon recent results from these conflicting fields.

In this article we focus on evaluation considering results from communication studies, a very broad field of social sciences. Evaluation goes beyond the statement of how much, but rather to concern itself with the question of "what value". It seeks to cope with the tutor and student questioning of "what progress am I making". Evaluation, therefore, presupposes a definition of goals to be reached – objectives that have been set forth [8].

Furthermore, recalling Suchman's statement that "how people work is one the most kept secrets of America" [11], we suggest that evaluation is by many means a silent communication between teacher-student, sons and parents, peers, and institutional settings. If we accept that evaluation can be viewed as set of invisible rules known by nobody, but somehow understood by everybody, we should add anthropological inquiries to our research toolkit.

The paper is organized around some elements of the social communication and how they convey new insights to evaluation research for computer and related scientists. We point out some technical limitations and offer discussions on how we relate to technology at same time we establish expectancies and perceive others work.

## 2. COMMUNICATION MODELS

Communication has been firstly and successfully modeled as an emitter and a transmitter connected by a channel [10]. This model rose from the needs of telegraphic communication, such as the range of frequencies for a physical channel, or in other words, the amount of information that may be conveyed from emitter to transmitter. It enabled the study of issues such as buffering strategies, time division and frequency division multiplexing, error detection, self-recovery encoding, and cryptography.

Many advances were built over the telegraphic communication model. If we reflect upon the character of such advances, we shall notice that social sciences probably have trouble to address its investigations with the telegraphic model. The reason is the dramatic simplification it inflicts to the concept of "communication", which could simply be interchanged with "transmission". The point is, although such model was very successful in many areas, for social sciences a new model needed to be crafted, as argued by Winkin [14]. We build on that to inquire whether evaluation research was also rolling on the same inadequate track.

We suggest that evaluation has a strong communication aspect and investigate it from the orchestral model perspective. In the orchestral model, communication is depicted as a permanent social process that integrates multiple behaviors: speech, gestures, looks, mimics, and space between people. Evaluation as communication would comprise all these attributes.

## 3. ELEMENTS OF THE ORCHESTRAL COMMUNICATION

In this section, we selected five research findings from anthropologists, psychiatrists, and social scientists to describe the orchestral communication model. For each of these findings we discuss how they construct the notion of evaluation as communication.

**Double Bind**
The double bind consists of a contradictory relation to whom one person stimulates the other just up to the point the other starts responding. The contradiction stems from ceasing the stimuli as soon as the response starts. Individuals get trapped in a paradoxical situation, because they do not know whether to be responsive or to be indifferent. Bateson *et al.* [4] first realized it while studying the Balinese mother and her child. In later studies about schizophrenia, Bateson and colleagues crafted the concept of double bind. They argued that the disease was deep rooted in the net of contradictory relations held within a human family. Bateson broadened the double bind concept to find it in religion, humor, and other social behaviors. In this way, double bind is a finding that supports many of the work in orchestral communication, especially because it objectively relocates communication from the individual to the quality of the relation kept among individuals.

Evaluation is also trapped in the double bind. Student and teacher or even the other characters such as colleagues and parents exchange many contradictory stimuli about learning. For example: have critical sense versus accept as truth what is in books, express yourself efficiently versus do not talk, concentrate on homework versus play with friends, etc. These situations are inherent to evaluation in the same sense the double bind is part of communication.

**Problem Reframing**
Problem framing consists of a two-phase communication intended to resolve paradoxical situations. Such situations characterize themselves as a dilemma in which no action is possible. For example, Watzlawick *et al.* [13] present a situation in which a police officer is in charge of evacuating a park during a riot: "Mr and Mrs, I have orders to shoot over the bandits. But I see in front of me many honest and respectable citizens. I request you to leave in order that I can shoot the bandits without more risks." This example demonstrates the two-phase communication for the problem. First, the officer makes clear his objective in a way he restricts possibilities. Second, the officer enables the construction of a silent agreement by reframing the problem. The problem is not anymore suppressing the riot; it is shooting the bandits without hurting the honest citizens.

Teachers do the same problem reframing while dealing with evaluation. Consider for example the following dialog with a group of students waiting to start an examination:

**First phase:** The test is to help me (teacher) to know how far you went in learning this subject; therefore cheating will be severely punished.
**Second phase:** Since there is nothing in the test we did not see in our classes, you assiduous and committed students will do very well.

The objective is clear here; get students and teacher out of a dilemma. That is, make students engage calmly in a difficult final exam without cheating. What clearly happens is also a problem reframing. It is not anymore threatening students not to cheat. It is considering them all obvious good students, which will do a good exam.

**Vestiges of mother language communication while expressing in a foreign language**
In an observation of Canadian natives of the Kutenai tribe, Birdwhistell [5] realized changes in the gestures while natives switch from Kutenai to English language.

The conclusion is that words are accompanied by the gestures and it seems to be inherent to communication.

Taking this standpoint, evaluation aspects such as clarity, consistency, and expressivity not necessarily traverse the many subject contents a student works on and is examined for. Evaluation as communication is a way to rescue content (which was always so badly dismissed as second class knowledge by interactionists) as an important reference for evaluation. Learning is not restricted to content assimilation, but content plays an important role in how we perform our knowledge (be evaluated). It seems obvious, but only because (parodying Macllulan) "evaluation is the message".

**Meta-Communication**
In a study in the zoo of San Francisco, Bateson [3] described how two lontras interact. He observed these animals playing with a piece of paper or fish he intentionally dropped. Apparently, the animals perform a real battle for the toy, but they do not hurt themselves. They know it is a game. With sounds and movements they qualify their interaction, their communication. In other words, they meta-communicate, that is, they communicate about their communication. Bateson used this conclusion to clarify the double bind concept. When the mother stimulates the child, the child responds, the mother ceases the stimuli, the child stops responding. Then the mother resumes the stimuli again. This action is a *commentary* on the action of ceasing the stimuli and can be understand as a meta-communication.

Similarly, evaluation is also a meta-communication in many situations. The situation in which the teacher asks for attention by saying "this will be asked in the exam" consists of a commentary on the loudly "please pay attention" request. Another example is the use of signs to classify the level of difficulty of book exercises. Such signs describe how the communication will occur when the student try to solve the problem. We can interpret these signs as a meta-evaluation, meaning that the solution of a difficult problem requires higher skills. So, the challenge could encourage the student to work on these exercises. On the other hand, there are many students that state beforehand that they will not be able to solve such problems and do not even read them.

**You cannot "not communicate". Nothing "never happens". The Silence.**
Communication is the performance of culture [9]. By culture we take everything you need to know in order to be part of, or to understand, the internal working of a social group. In terms of communication, it is a set of rules almost nobody knows, but almost everybody understands. Silence is the space for possibilities of communication, building meaning, expressing disapproval or for common agreement. The silence is a line over which the discourse punctuates it with words [7].

Evaluation needs to cope with silence. A blank answer is not without a meaning and depends on facts outside computer-based evaluation. Some meanings could be: there was not enough time to answer all questions, the student did not have a clue of what to develop there, the student was ashamed of telling what he/she thought would be the answer, the student was bored and simply did not take the effort to develop an answer. These are just a few possible meanings for the silent evaluation.

## 4. CONSEQUENCES OF EVALUATION AS COMMUNICATION

**Explicitation and Expression of Contradictory Views**
The hypothesis of investigation is that such communication framework could unveil new evaluation concepts. In respect to methodology, we suggest the design of candidate concepts in a computer-supported environment as a mean to clarify details and validate their integration with current practice.

Would it be possible to have a technology-enhanced evaluation to be a *performance of knowing*, in the same way Scheflen [9] considers communication as the performance of culture? For example, how a person regards on a theory to build a "sound argument" or how pilots regard on their flying skills to provide a "smooth landing". Performance of "sound argument" or "smooth landing" is a totalizing word for the perception and the reasoning of an audience.

Since we still do not have such new elements, we now provide a clue of what a computer-based learning system should provide.

In realistic situations subjects sometimes fail to solve problems that they fully understand, or provide amazingly correct responses to states that are not directly observable. The usual way out of such difficulties is a probabilistic approach as suggested by Doignon and Falmagne [6]. We are not concerned with explaining how evaluation works and how it should work. Our point is to know the system requirements that would enable the concept of evaluation as communication. Next we suggest some general lines for evaluation scenarios.

One of the first aspects of computer-based learning environments that must be adapted and extended is the way problems are proposed, the traditional "I-put-the-question-you-give-me-the-answer". Why could not a student's question be a valid form of answering to a problem? Actually, the solution may be the question and then you answer with another solution, such as a kind of reasoning by demonstration or deduction.

Another important feature is to account for and to explicitate the dimension of silence. Instead of simply recording student's activity, the environment should also point out which interactions he/she chose not to participate in. Reaction to meta-communication hints and the use of interactive corrections and problem solving tools by the students, without the pressure of an examination or the barrier of being criticized by another human that would read the answer, are also objects of interest.

Evaluation in such systems should not be perceived as defined by a single person perspective, but must be the resulting expression of multiple voices. In this aspect, the system should support a form of meta-evaluation. For example, it could take into account what the students think about evaluation and in which aspects they judge they should be evaluated. It should be able to collect and confront the perceptions of one self-performance with others perceptions of him or her. Some ways to support this broader view are to expand one's audience, by selecting or reinforcing public access for specific exercises; to select one's audience with and without student or teacher knowing; and to support the exchange of comments on each others answers.

To enrich this evaluation experience, the computer-based learning environment must support and, if possible, amplify the expression and the emergence of contradictory relations. These are essential to the evaluation process, since provocative statements may communicate how teacher and colleagues perceive one's performance and understanding. One of the ways to promote such type of relations is to select different audiences to appreciate a specific exercise solution. It would also be interesting for the environment to be able to refer to some relevant communication that happened out of the system – for example, to refer to corridor conversations and to content outside the current subject.

Thus, to develop a system to achieve the desired performance of knowing, there is an inherent need for providing evaluation information for a broader audience, rather than just the teacher, and to diverse the composition of evaluation and assessment elements, such as written and oral tests, opinions of others, recordings of participations in class, and so on.

All of the above points could be inspiring to conceive novel computer-based evaluation scenarios, but since conflicting views emerge every time we deal with evaluation, there still need to be long rounds of discussion, negotiation, and commitment. The next section elaborates more on it by means of a succinct case study in the context of a real project.

**Conciliating Conflicting Views: A Case Study**
We developed an annotation tool as part of a computer-based educational system, named CALM [1]. We had an evolutionary approach of adding new functionalities and exploring new uses for the annotation technology. We envisaged three scenarios, in increasing level of complexity: on-line asynchronous discussion, collaborative text authoring, and collaborative evaluation. The first and simpler was the support for discussion over a prepared hypertext available in the Web. The idea was to have students annotating the document and discussing about it. Annotations would be provided with symbols to denote agreement, disagreement, new facts, and issues. The second scenario was to support collaborative document edition by means of annotating and incorporating annotations to the final document. It made use of a more complex workflow for voting and consolidating annotated material. The last scenario, which is again an evolution of the former ones, was to have a teacher and a student building an evaluation by means of exchanging annotations on the answers of an exam. The student would be doing the exam and exchanging points by receiving and doing annotations with the teacher. All these scenarios adopted three metaphors of annotation, demonstrated in Figure 1, which are explained in detail by Adriano *et al.* [2].

The evaluation scenario inspired the discussion to have evaluation as a form of communication. Our aim was to use the annotations to reify this idea. Our approach of evolving technology and corresponding use led us to many obstacles. We ignored the differences of conceptualization that educators and engineers may have for "annotation", "discussion", "authoring", and, more dramatically, "evaluation". Three aspects were central:

ergonomics, economics, and politics. Ergonomics in respect to issues such as where to place annotations, when annotate, visibility of annotations. Economics, in respect to which browsers to support and which XML-schema to adopt. Politics in respect to the kind of content we should provide in CALM and in which groups or classes to probe the scenarios.

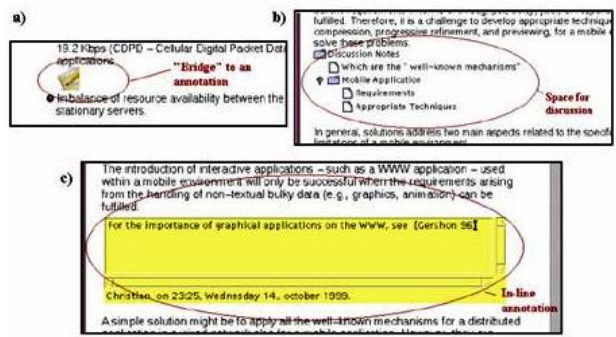

*Figure 1.* Three annotation scenarios in CALM.

At this point, we suggest that none of the previous aspects would have been neglected if we departed from an anthropological analogy of our contending forces. The benefit of such an anthropological background is to obtain consensus by having ideas moving forward in a common ground of shared and agreed concepts. The direct consequence is the diminishment of the internal race for one or other view of technology and its appliances.

**5. CONCLUSION**

What kind of light such a study brings for researchers developing interactive media? Evaluation as a communication, as approached here, is a standpoint to understand how epistemologies such as educational technology adopt innovation (especially technical enhanced ones). From the technophile to the minded educator, there is a large range of possible situations to adapt to technology and to have it adapted to evaluation purposes. Issues of economy, ergonomics, and politics must all have their turn to influence engineers and educators decisions.

Moreover, by promoting evaluation to an instance of communication, we enabled the appliance an anthropological toolset for our research. The novelty is to suggest new ways of combining technology with evaluation aims. The utopia would to be to rescue evaluation from the school and the academy ghettos. This comprise many statements, that could be put in the form of a manifest as follows: (1) Student knowledge should perform for ample audiences; (2) Evaluation should be a truly interactive and collaborative experience; (3) Evaluation as lifelong and social pervasive practice; (4) Evaluation as stakeholder for important content assimilation; (5) Evaluation to promote creativity, critical sense, and disposition to take risks.